\documentclass[letterpaper, 10 pt, conference]{ieeeconf} 
\usepackage{amsmath} 
\usepackage{amsfonts} 
\usepackage{amssymb}
\usepackage{trfsigns}
\usepackage{mathrsfs} 
\usepackage{graphicx} 
\usepackage{xcolor}
\usepackage[]{units}
\usepackage[]{xspace}

\usepackage{booktabs}							

\IEEEoverridecommandlockouts 
\newcommand*{\red}[1]{\textcolor{black}{#1}}
\newcommand*{\green}[1]{\textcolor[rgb]{0,0,0}{#1}}

\newcommand*{\ie}{i.e.,\xspace}

\newcommand*{\eg}{e.g.\xspace}

\newcommand*{\abs}[1]{\lvert#1\rvert}			
\DeclareMathOperator*{\argmin}{arg\,min}

\newcommand*{\diff}{\mathop{}\!\mathrm{d}}	

\newcommand*{\DV}{\ensuremath{\Delta V}\xspace}
\newcommand*{\Pot}{\ensuremath{\Phi_\text s}\xspace}
\newcommand*{\Pots}{\ensuremath{\Phi^*}\xspace}
\newcommand*{\Tscan}{\ensuremath{T_\text{scan}}\xspace}
\newcommand*{\Vneg}{\ensuremath{V^-}\xspace}
\newcommand*{\Vpos}{\ensuremath{V^+}\xspace}
\newcommand*{\Vmp}{\ensuremath{V^\mp}\xspace}
\newcommand*{\Vmpp}{\ensuremath{V^\mp(p)}\xspace}
\newcommand*{\Vb}{\ensuremath{V_\text{b}}\xspace}
\newcommand*{\Vbmod}{\ensuremath{V_\text{b,mod}}\xspace}
\newcommand*{\Vbc}{\ensuremath{V_\text{b,C}}\xspace}
\newcommand*{\Vbff}{\ensuremath{V_\text{b,FF}}\xspace}
\newcommand*{\Df}{\ensuremath{\Delta f}\xspace}
\newcommand*{\Dfref}{\ensuremath{\Df_\text{ref}}\xspace}
\newcommand*{\Dfdref}{\ensuremath{\Df'_\text{ref}}\xspace}
\newcommand*{\ef}{\ensuremath{e_f}\xspace}
\newcommand*{\ed}{\ensuremath{e_d}\xspace}
\newcommand*{\eSTC}{\ensuremath{e_\text{STC}}\xspace}
\newcommand*{\ELP}{electrostatic potential\xspace}
\newcommand*{\ELPs}{electrostatic potentials\xspace}
\newcommand*{\PLL}{phase-locked loop\xspace}
\newcommand*{\WPLL}{\ensuremath{\omega_\text{PLL}}\xspace}
\newcommand*{\FF}{feedforward\xspace}
\newcommand*{\SQDM}{scanning quantum dot microscopy\xspace}
\newcommand*{\Kesc}{\ensuremath{K_\text{ESC}}\xspace}
\newcommand*{\Kstc}{\ensuremath{K_\text{STC}}\xspace}
\newcommand*{\Ad}{\ensuremath{a_d}\xspace}
\newcommand*{\Wd}{\ensuremath{\omega_d}\xspace}
\newcommand*{\WL}{\ensuremath{\omega_\text L}\xspace}
\newcommand*{\WH}{\ensuremath{\omega_\text H}\xspace}

\title{\LARGE Control of Scanning Quantum Dot Microscopy}
\author{
  Michael Maiworm$^{1}$, 
  Christian Wagner$^{2}$, 
  Taner Esat$^{2}$,
  Philipp Leinen$^{2}$,
  Ruslan Temirov$^{2}$,\\ 
  F. Stefan Tautz$^{2}$, 
  Rolf Findeisen$^{1}$ 
  \thanks{$^{1}$Otto-von-Guericke-Universit{\"a}t Magdeburg,
    Laboratory for Systems Theory and Automatic Control, Germany,
    $\lbrace$rolf.findeisen,
    michael.maiworm$\rbrace$@ovgu.de. 
    $^{2}$Peter Gruenberg Institute (PGI-3), Juelich Research  Center,
    Germany, c.wagner@fz-juelich.de.}
  }

\begin{document}
\maketitle
\thispagestyle{empty} \pagestyle{empty}

\begin{abstract}
  Scanning quantum dot microscopy is a recently developed
  high-resolution microscopy technique
  that is based on atomic force microscopy and is capable of imaging
  the electrostatic potential of nanostructures like molecules or
  single atoms.
  Recently, it could be shown that it not only yields qualitatively but
  also quantitatively cutting edge images even on an atomic level.
  In this paper we present how control is a key enabling element to
  this.
  The developed control approach consists of a two-degree-of-freedom
  control framework that comprises a feedforward and a feedback part.
  For the latter we design two tailored feedback controllers. 
  The feedforward part generates a reference for the current scanned
  line based on the previously scanned one.
  We discuss in detail various aspects of the presented control
  approach and its implications for \SQDM.
  We evaluate the influence of the feedforward part and compare the two
  proposed feedback controllers.
  %
  The proposed control algorithms speed up \SQDM by more than a magnitude
  and enable to scan large sample areas.
\end{abstract}


\section{Introduction}
\label{sec:introduction}

An important aspect in nanotechnology is the determination of
characteristics of the fundamental building blocks of matter, namely
atoms and molecules.
One of these characteristics are the electrostatic properties that
govern in many cases the functionality of nanoscale objects and systems.
This is particularly important for new materials and devices associated
with nanoscale electronics, such as semiconductors.
The investigation of electrostatics at the nanoscale becomes therefore
more and more important and is a vivid field of ongoing research
\cite{Matyba2009, Musumeci2014, Fuchs2016, Hu2016, Wagner2019a}.

\begin{figure}[tb]
  \centering
  \includegraphics[width=1.0\linewidth]{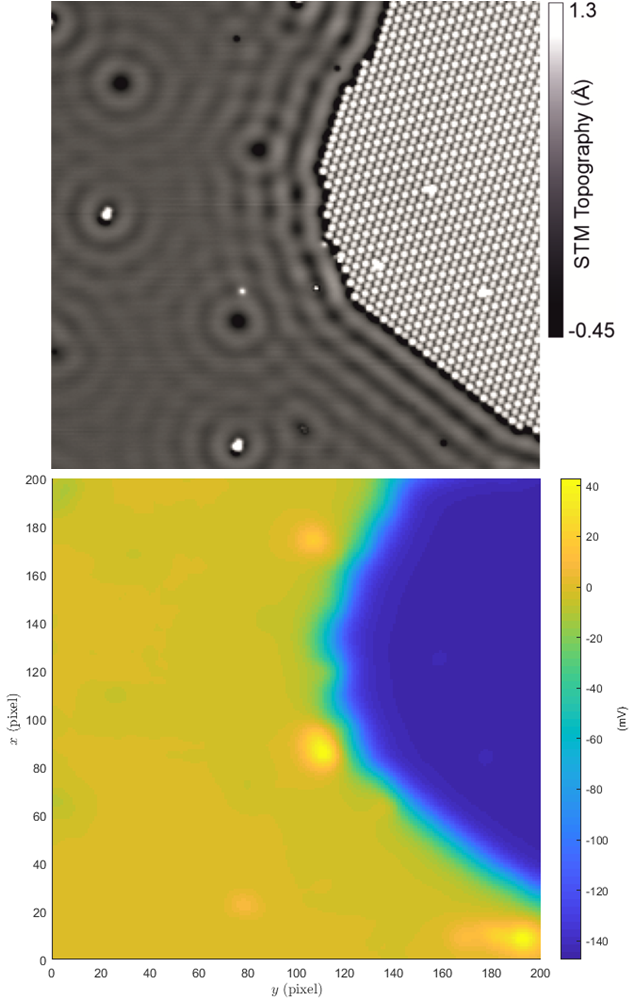}
  \caption{
    Top: Scanning tunneling microscope image of a sample presented in
    \cite{Wagner2019a}.
    The image size is \unit[600$\times$600]{\AA}.
    Bottom: 2D \ELP image of the same sample at height $z =
    \unit[20]{\AA}$.
    The image was divided into 200$\times$200 pixels.
  }
  \label{fig:STM_SQDM}
\end{figure}

Scanning quantum dot microscopy (SQDM), introduced in
\cite{Wagner2015, Green2016} allows to measure the \ELPs of
nanostructures with sub-nanometer resolution at large imaging
distances.
It generates 2D images of the \ELP (Fig.~\ref{fig:STM_SQDM}).
It furthermore allows to separately image the electrostatic potential
and the surface topography.
Recently it was shown in \cite{Wagner2019a} that SQDM does
not only generate qualitative but also quantitative images of the
electrostatic potential of nanostructures.
The paper also demonstrated large-scale imaging
(Fig.~\ref{fig:STM_SQDM}), resolving both small (single atoms and
molecules) and large structures (an island composed of several hundreds
of molecules) in the same image, whereas in the previous publications of
SQDM \cite{Wagner2015, Green2016} only isolated atoms and molecules were
imaged independently of each other.
Thereby, SQDM became a mature widely applicable microscopy technique for
the area of nanotechnology.

This paper is the control engineering counterpart of \cite{Wagner2019a},
wherein the controller was only briefly mentioned.
We present in detail the tailored two-degree-of-freedom
control algorithm that is a key enabling element that turned SQDM into
a well applicable microscopy technique.
The controller is a variant of the one presented in \cite{Maiworm2018},
which included a Gaussian process as a feedforward signal generator.
However, that approach is not yet ready for the deployment in the
experiment because further research, in particular regarding the
computational issues for the online learning of the Gaussian process,
has to be conducted.
Henceforth, the objective of this work is to present the controller
version that was actually used in \cite{Wagner2019a}.
Opposed to the initial method of generating SQDM images based on
spectroscopy grids (see \cite{Wagner2015}), the controller now allows to
continuously scan the sample, which yields order-of-magnitude faster
image generation and eliminates the need for spectroscopy.
This puts it in line with other microscopy techniques like
scanning tunneling microscopy \cite{Binnig1982} and atomic force
microscopy \cite{Albrecht1991}.
Furthermore, the controller now also allows to scan images with highly
varying \ELPs.
Previously, using spectroscopy grids one had to set up the spectrum
range, which increases with the \ELP, before starting the grid. 
This was a huge setback in speed if the \ELP and with that the spectrum
range was large because the entire voltage range had to be scanned.

~\\
After this introduction, the paper is structured as follows.
Sec.~\ref{sec:scanning_quantum_dot_microscopy} provides the
fundamentals of \SQDM.
The 2DOF control framework is presented in Sec.~\ref{sec:control} and
thoroughly analyzed in simulations in Sec.~\ref{sec:simulations}.
Further experimental results are presented in
Sec.~\ref{sec:experimental_results} and the paper is concluded in
Sec.~\ref{sec:conclusion}.

\section{Scanning Quantum Dot Microscopy}
\label{sec:scanning_quantum_dot_microscopy}

This section explains the working principle of \SQDM and the associated
image generation process. The process is analyzed and a model derived
that will be used in Sec.~\ref{sec:simulation_results} for simulations.

\subsection{Working Principle}
\label{sec:working_principle}

\begin{figure}[tb]
  \centering
  \includegraphics[width=0.9\linewidth]{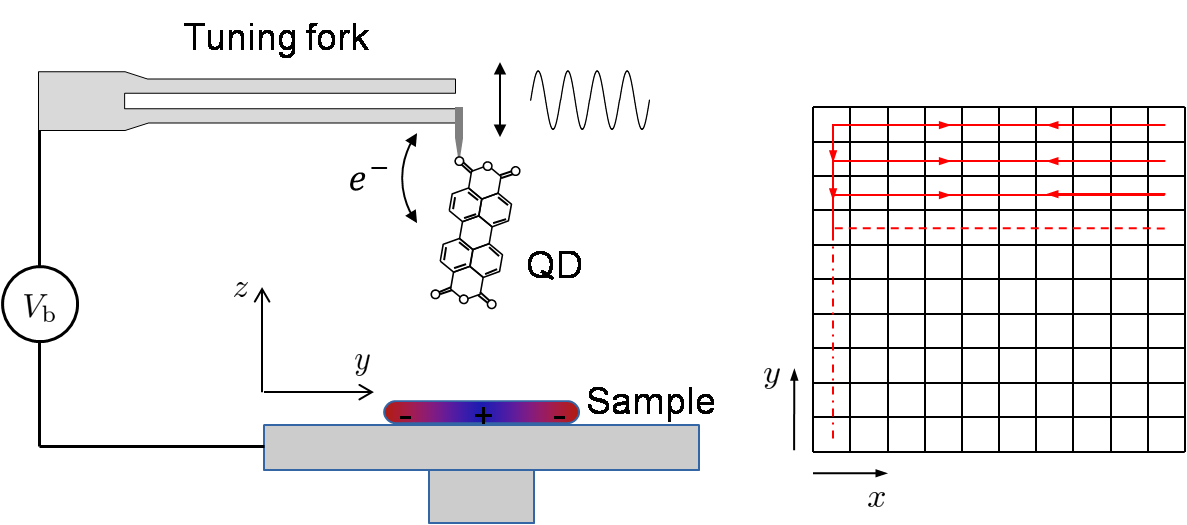}
  \caption{Schematic of Scanning Quantum Dot Microscopy (left). The tip
  of a frequency modulated non-contact AFM is decorated with a quantum
  dot (QD) and a bias voltage source is connected between the microscope
  tip and the sample. 
  Depending on the \ELP of the nanostructure on the sample surface and the
  bias voltage \Vb, a single electron ($e^-$) tunnels back and forth
  between the AFM tip and the quantum dot.
  The tip together with the quantum dot is moved in a raster scanning
  pattern (right).}
  \label{fig:SQDM_schematic}
\end{figure}

SQDM is able to measure electric surface potentials and allows to
distinguish between topographical and electrostatic effects.
It utilizes a frequency modulated non-contact atomic force microscope
(NC-AFM) \cite{Albrecht1991}, operated in ultra-high
vacuum and at a temperature of \unit[5]{K}.  
The atomically sharp tip (Fig.~\ref{fig:SQDM_schematic}) is mounted to a
tuning fork (qPlus sensor \cite{Giessibl2003}) that oscillates with a
frequency $f = f_0 + \Df$ of around \unit[30]{kHz}, where
$f_0$ is the free resonance frequency and \Df is the frequency change that
is caused by a vertical force gradient acting on the tip.
Typically, this force is the result of the tip-sample interaction.

The AFM tip is decorated (\cite{Toher2011, Fournier2011, Wagner2012,
Wagner2014, Findeisen2016}) with a quantum dot (QD)\footnote{Currently a
PTCDA (Perylenetetracarboxylic dianhydride) molecule serves as the QD.},
a nano-sized object whose energy levels can take only discrete values.
Changes of the \ELP \Pot of the surface can change the
QD's charge state via gating as an electron tunnels from the tip into
the QD.
This leads to an abrupt change in the tip-sample force.
These tip-sample force changes are detected by the NC-AFM,
effectively transducing the information about the electrostatic
potential of, \eg a nanostructure, into the measurable quantity \Df. 
Monitoring the charging events of the QD while scanning the sample is
the basic working principle of SQDM. 

To detect the charging events, a bias voltage source \Vb is
connected to the sample while the tip is grounded.
The associated \ELP $\Phi_\text{b}$ that is generated by \Vb is
superimposed on the intrinsic electrostatic surface potential of the
sample \Pot.
Accordingly, a change in \Vb then leads to a change of the effective
electric potential at the QD.  
If this reaches a threshold value, a change in the QD's charge state
is triggered.

\begin{figure}[tb]
  \centering
  \includegraphics[width=\linewidth]{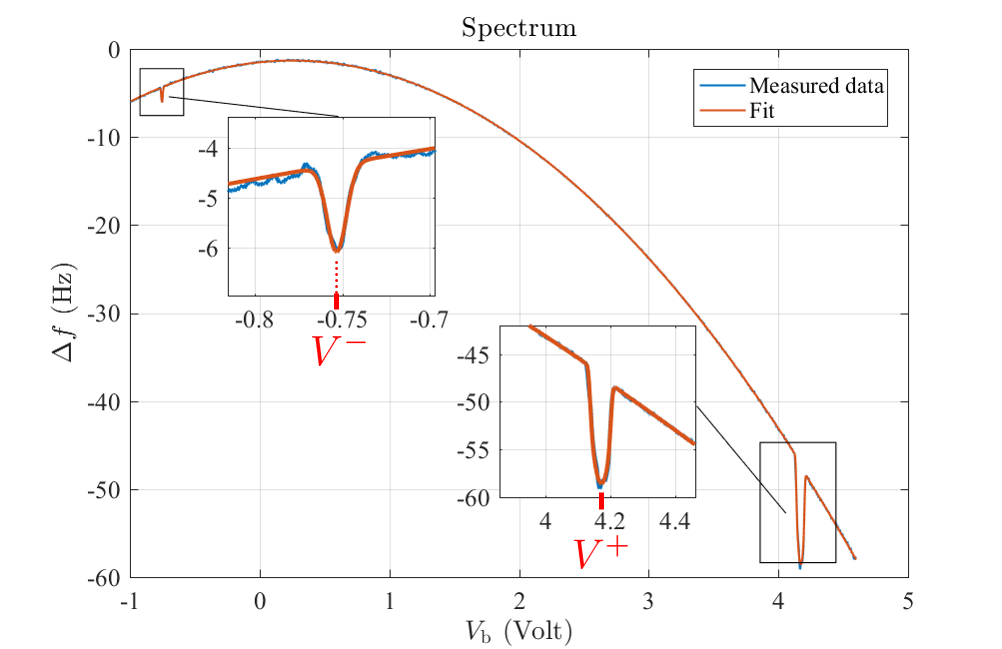}
  \caption{The spectrum $\Df(\Vb)$ describes how the tuning fork
  oscillation frequency changes with the bias voltage \Vb at one
  particular position in space. 
  The overall parabolic shape of the curve is a consequence of the
  tip-sample capacitance (see \cite{Gross2009}).
  The SQDM specific dips result from the charging events of the quantum
  dot.
  The voltage values where the two dips reach their minimum are
  indicated by \Vneg and \Vpos.
  }
  \label{fig:spectrum}
\end{figure}

Charging of the QD leads to a change in the tip-sample force, whose
gradient is proportional to \Df for small amplitudes of the AFM tip
oscillation (see \cite{Giessibl1997}).
The changes of the tip-sample force generated by the charging events
appear in the so-called \emph{spectrum} $\Df(\Vb)$ as
features that we denote as \emph{dips} (Fig.~\ref{fig:spectrum}).
The $\Df(\Vb)$ spectrum is the superposition of a parabola (\cite{Gross2009})
and two dips, one at negative \Vb values and one at positive \Vb values.
The dips separate \Vb intervals with different charge states of the QD.


We denote the voltage values at which the dips reach their minimum with
\Vneg and \Vpos, or short \Vmp.
These values  characterize the dips' positions within the spectrum.
By means of
\begin{align}
  \Pots(p) = \frac{\Vneg_0 \cdot \DV(p)}{\DV_0} - \Vneg(p) \ ,
\label{eq:SQDM}
\end{align}
the effective surface potential \Pots at the position of the tip $p = (x,y,z)$
can be calculated, where $\DV = \Vpos - \Vneg$ and $\Vneg_0, \DV_0$ are
reference points.
The actual surface potential \Pot can be recovered from \Pots
through deconvolution in post-processing \cite{Wagner2019b}.
In the rest of this paper we will deal with \Pots unless otherwise
indicated.
For more details see \cite{Wagner2019b}.

\begin{figure}[htpb]
  \centering
  \includegraphics[width=\linewidth]{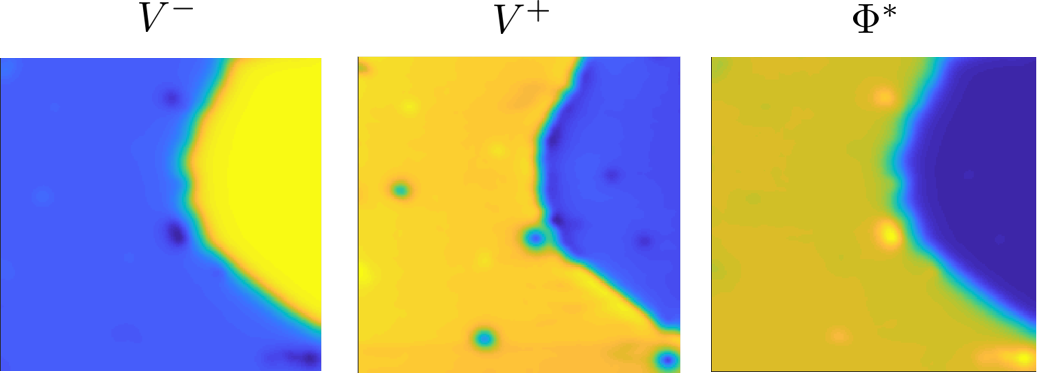}
  \caption{
  From left to right: the \Vneg map, the \Vpos map, and the
  resulting electrostatic potential image \Pots of
  Fig.~\ref{fig:STM_SQDM}.
  }
  \label{fig:dipMaps}
\end{figure}

The \Vmpp maps together with $\Pots(p)$ of the sample shown in
Fig.~\ref{fig:STM_SQDM} are depicted in Fig.~\ref{fig:dipMaps}.

\subsection{Original Image Generation Process}
\label{sec:original_image_generation_process}

The image generation process that had been previously used in
\cite{Wagner2015, Green2016} is as follows.
The sample (Fig.~\ref{fig:STM_SQDM}) is discretized in pixels
(Fig.~\ref{fig:SQDM_schematic}) and the tip with the QD is moved from
pixel to pixel.  
At the first pixel, a complete spectrum (like Fig.~\ref{fig:spectrum})
is measured and the positions of the dips \Vmp are determined.
It has to be assumed that the interval in which \Vmpp will change while
scanning the sample is approximately known a priori. 
For the following pixels, the bias voltage \Vb is swept accordingly
within these two intervals (\eg a voltage range of \unit[0.2]{V} instead
of \unit[6]{V} for the complete spectrum). 
This results in the measurement of local dip spectra. 
After obtaining the local dip spectra for all pixels, the \Vmpp
values are determined for each pixel and used in \eqref{eq:SQDM}
to generate \Pots.

The main limitation of this image generation process is the required
large total measurement time.
For instance, measuring the local dip spectra takes about \unit[3]{s} for
each dip and pixel for a certain \Vb interval size.
Hence, the determination of the complete \Vmp maps in
Fig.~\ref{fig:dipMaps} would require \unit[66.7]{h}.\footnote{The data
shown in Fig.~\ref{fig:dipMaps} was generated by the proposed control
approach in this paper, which results in significantly smaller
measurement times.}
This severely limits the applicability of SQDM.
In particular, 
\begin{itemize}
  \item the microscope is blocked for several hours for the generation
    of one image,
  \item the longer the measurement time, the higher the probability of
    failures, and
  \item effects like drift increase and deteriorate image quality.
\end{itemize}
Furthermore, obtaining \Pots images from a grid of spectra limits SQDM
in several ways:
\begin{itemize}
  \item Generation of fast and rough images for a first impression is
    not possible.
  \item Generation of images like the one presented in
    Fig.~\ref{fig:STM_SQDM} and Fig.~\ref{fig:dipMaps}
    become practically impossible.
  \item The measurement times of the local spectra increase if the
    aforementioned voltage intervals, wherein the dips move, increase.
    This can occur, for instance, for other substrate-sample
    combinations with stronger electrostatic variations and thus, reduce
    the applicability even further.
\end{itemize}

\subsection{Simulation Model}
\label{sec:simulation_model}

The controllers, as outlined in Section~\ref{sec:control}, are going to
be evaluated in simulations in Section~\ref{sec:simulations}.
This requires a SQDM simulation model, which we derive in the following.

\begin{figure}[htpb]
  \centering
  \includegraphics[width=0.7\linewidth]{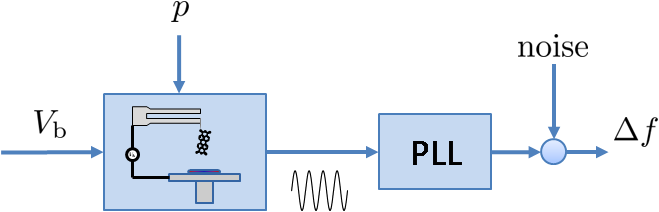}
  \caption{SQDM block diagram.
  }
  \label{fig:block_diagram_SQDM}
\end{figure}

Fig.~\ref{fig:block_diagram_SQDM} illustrates the SQDM system in a block
diagram.  
In the experiment, the current bias voltage \Vb and tip
position $p$ lead to an oscillation frequency $f$ of the tuning fork
different from the free resonance frequency $f_0$. 
Hence, \Vb and $p$ are the inputs to the system and the frequency change
\Df is the output.
The bias voltage \Vb is the control input to the system and
can be chosen freely, whereas the tip position $p$ changes according
to the raster scanning pattern as shown in Fig.~\ref{fig:SQDM_schematic}.
As \Vb changes, the frequency changes almost instantaneously according
to Fig.~\ref{fig:spectrum}.  
Therefore, the first block in Fig.~\ref{fig:block_diagram_SQDM} can be
considered stationary.  
The relatively small frequency change \Df (only up to a few Hz,
compared to \unit[30]{kHz} of the free resonance frequency
$f_0$) is determined from the oscillation signal using a \PLL (PLL),
which is modeled as a first order system with bandwidth
\WPLL and whose output is corrupted by white Gaussian noise.

To simulate SQDM and the image generation process, the spectrum 
$\Df(\Vb)$ has to be available as an analytic function.
It consists of a parabola and the two dips, which we model as
Gaussian curves\footnote{
  Note that in the actual experiment, the shape of the dips is somewhere
  between a Gaussian curve and a half-circle, depending on the chosen tip
  oscillation amplitude, the value of \Vmp, and the width of the
  electronic level of the QD.
  See also \cite{Kocic2015}.
}.
The ansatz for the spectrum is 
\begin{align}
  \Df(\Vb) &= \Df_\text{para}(\Vb) + \Df^-(\Vb) + \Df^+(\Vb)
\label{eq:spectrum_fit}
\end{align}
with the parabola function
\begin{align*}
  \Df_\text{para}(\Vb) = p_1 \Vb^2 + p_2 \Vb + p_3
\end{align*}
and the Gaussian curves
\begin{align}
  \Df^-(\Vb) &= d^- \cdot
      \exp\!\left(
        -\!\left(\frac{\Vb-\Vneg}{w^-}\right)^2
          \right)  \label{eq:negPeak}  \\
  \Df^+(\Vb) &= d^+ \cdot
      \exp\!\left(
        -g\!\left(\frac{\Vb-\Vpos}{w^+}\right)
          \right)  \label{eq:posPeak}
\end{align}
for the dips, where $d^\mp, \Vmp, w^\mp$ are the respective depth,
position, and width of the dips.
The function $g(\cdot)$ in \eqref{eq:posPeak} is
the polynomial $g(x) = a_1 x^2 + a_2 x^4 + a_3 x^6$.
The parameters in \eqref{eq:spectrum_fit}, \eqref{eq:negPeak}, and
\eqref{eq:posPeak} are then fitted using experimental data. 
The resulting fit in Fig.~\ref{fig:spectrum} and
Tab.~\ref{tab:spectrumFit} shows that the proposed
ansatz is well suited to model the experimentally acquired \Df spectrum.

\begin{table}[htpb]
  \centering
  \caption{\Df Spectrum Fit Parameters}
    \begin{tabular}{llll}
    \toprule
      $c_1 = -1.3$ & $a_1 = 0.70$  & $d^- = -1.1$ & $d^+ = -4.6$ \\
      $c_2 = 0.56$ & $a_2 = -0.61$ & $V^- = -1.3$ & $V^+ = 4.3$ \\
      $c_3 = -0.76$ & $a_3 = 1.64$ & $w^- = 0.022$ & $w^+ = 0.087$ \\
    \bottomrule
    \end{tabular}
  \label{tab:spectrumFit}
\end{table}

To simulate the changing tip position, the experimental reference
data of Fig.~\ref{fig:dipMaps} for \Vmp for every pixel
is fed to \eqref{eq:negPeak} and \eqref{eq:posPeak}.  
Thus, at each new pixel in the simulation, the dips are
shifted according to the experimental reference.

\section{Controller Development}
\label{sec:control}





The effective \ELP of the sample at the position of the tip $\Pots(p)$
changes in the three dimensional space.
However, since SQDM is based on AFM, only 2D images can be generated.
Therefore, we consider in the following $\Pots(x,y)$ that, according to 
\eqref{eq:SQDM}, depends in turn on the quantities $\Vmp(x,y)$, \ie
$\Pots\big( \Vmp(x,y) \big)$.
The \Vmp values are a priori unknown, cannot be measured directly, and
change with the tip position $p = (x,y)$.
Therefore, SQDM with the changing \Vmp values can be regarded as
a parameter varying system and the objective of this paper is to design
a control framework that automatically determines and tracks the unknown
parameters while scanning the sample.

As outlined, \Vmp cannot be measured directly and a model of the
respective dynamics is unavailable because it depends, besides the scan
speed, on the \ELP, which itself is unknown.
Thus, a potential control approach has to adapt \Vb indirectly, based on
a quantity that can be measured, in this case the frequency change \Df.
Hence, we are looking for a control law $\Vb(t) = \kappa(\Df)$.

\begin{figure}[htpb]
  \centering
  \includegraphics[width=\linewidth]{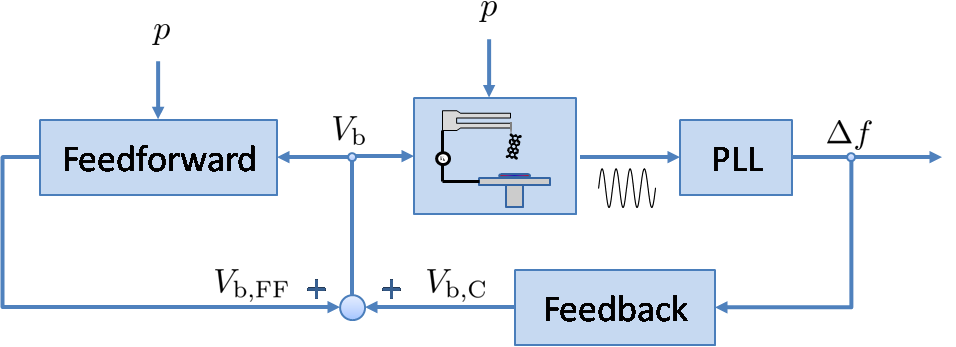}
  \caption{Block diagram of the closed loop: The bias voltage \Vb is the
  sum of the feedback and feedforward output. The feedforward part is
  influenced by the current tip position $p$.}
  \label{fig:block_diagram_general}
\end{figure}

In the following we present a two-degree-of-freedom (2DOF) control approach,
consisting of a feedback and a feedforward (FF) part
(Fig.~\ref{fig:block_diagram_general}).
We furthermore develop two different feedback controllers, leading to
two different versions of the 2DOF controller.
The first feedback controller is an extremum seeking controller (ESC)
and the second controller is called \emph{slope tracking controller}
(STC).
\green{The central idea to both controllers is, instead of measuring the dip
spectrum (time consuming, contains a lot of unnecessary data), to track
one specific reference point in each dip.
A major difference between the ESC and the STC is this reference point.
While the ESC tracks directly \Vmp, the STC tracks a point on the dip's
slope (thus the name).}

The individual 2DOF parts will be discussed in more detail in the remainder
of this section.

\subsection{Extremum Seeking Control}
\label{sec:extremum_seeking_control}

The fact that \Vmp characterize the minima of the local convex
dips $\Df^\mp$ satisfying
\begin{align}
  \Vneg(x,y) &= \argmin_{\Vb} \Df^-(x,y,\Vb)  \label{eq:Vneg} \\
  \Vpos(x,y) &= \argmin_{\Vb} \Df^+(x,y,\Vb) ,
    \label{eq:Vpos}
\end{align}
can be exploited for the determination of \Vmp.
Since \Df is measured online by the \PLL, the minima of $\Df^\mp$ can be
determined continuously by appropriately adapting the input \Vb in
\eqref{eq:Vneg} and \eqref{eq:Vpos} during the scanning process.

This can be achieved by employing methods of extremum
seeking control \cite{Ariyur2003, Zhang2012}.
As the name indicates, these methods are designed to find the extremum,
\ie a minimum or a maximum, of the output of a given system.
%
The core principle includes a seeking element that continuously samples
the output signal at the current operating point to obtain some kind of
direction or gradient information.
If a measure of the gradient is detected, the current operating point is
changed accordingly.
If the optimum is reached, the gradient is zero and the operating point
is not changed anymore.
Thus, ESC approaches are closely related to optimization and are
therefore also called \emph{real-time optimization} methods
\cite{Ariyur2003}.
In principle, an ESC can be realized with many different optimization
methods, though they have to satisfy a variety of additional properties,
such as, for instance, low computational load or the ability to deal
with constraints.

Works on extremum seeking date back to as early as 1922
\cite{Leblanc1922} or 1951 \cite{Draper1951}.
Though many other works were published in the second half of the 21th
century (see \cite{Guay2003}), it wasn't until the paper of
\cite{Krstic2000b} in the year
2000 that provided the first general stability analysis of ESC.
Since then, interest has sparked again and found its realizations in
applications such as anti-lock-breaking systems
\cite{Yu2002, Zhang2007}, maximum-power-point-tracking
\cite{Leyva2006, Brunton2010}, source seeking \cite{Zhang2007,
Cochran2009}, beam control in particle accelerators \cite{Scheinker2017},
or the control of plasma in a Tokamak reactor \cite{Centioli2008}.
Theoretic extensions for discrete time systems \cite{Choi2002},
for multivariable systems \cite{Rotea2000}, systems with partial model
information \cite{Guay2003, Adetola2006}, and further results on
stability \cite{Tan2006, Scheinker2017} have been presented.
Generalizations of the ESC scheme such as a unifying framework 
\cite{Nesic2013a} and with other optimization approaches, such as
newton like extremum seeking \cite{Moase2009, Ghaffari2012}, stochastic
\cite{Liu2012, Liu2014} and a non-gradient approach \cite{Nesic2013b}
have been developed too.
For extensive lists of further publications see \cite{Tan2010,
Zhang2012, Scheinker2017}.

\begin{figure}[htpb]
  \centering
  \includegraphics[width=\linewidth]{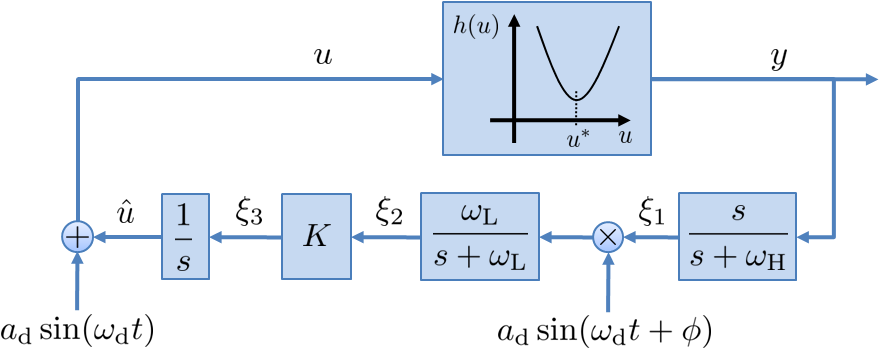}
  \caption{Block diagram of the extremum seeking control approach.}
  \label{fig:block_diagram_ESC_basic}
\end{figure}

In this work we employ an adapted version of the approach of
\cite{Krstic2000b} as shown in Fig.~\ref{fig:block_diagram_ESC_basic}.
In this setup we deal with a local convex function $h(u)$ for which
there exists a minimum at $u = u^*$.
At this minimum we have naturally $\frac{\diff h}{\diff u}
\big\rvert_{u^*} = 0$.
To search for $u^*$, a dither signal $d(t) = \Ad \sin(\Wd t)$ is used
to perturb the current $u$.
The resulting signal $y(t) = h(u(t))$ is passed through a high pass
filter $\frac{s}{s + \WH}$ to eliminate any constant offsets.
The filtered signal $\xi_1(t)$ is then multiplied by the phase shifted
dither signal\footnote{
  The phase of $\xi_1(t)$ is shifted by the high pass filter by $\phi$. 
  In order to be in phase, the dither signal that is multiplied with
  $\xi_1(t)$ is shifted also by $\phi$.} 
and low pass filtered with $\frac{\WL}{s + \WL}$.
This leads to $\xi_2(t)$ and it can be shown that
\begin{align*}
  \lim_{t \to \infty} \xi_2(t) = \frac{\Ad^2}{2} \frac{\diff h}{\diff u}
    \Big\rvert_{\hat u}
\end{align*}
at the current operating point $\hat u$.
Choosing $K := \nicefrac{-2}{\Ad^2}$, one obtains 
\begin{align*}
  \lim_{t \to \infty} \xi_3(t) = - \frac{\diff h}{\diff u}
     \Big\rvert_{\hat u}
\end{align*}
\ie $\xi_3(t)$ tends to the negative gradient of $h(\hat u)$.
Hence, $\xi_3(t)$ can be used subsequently to implement a gradient
descent approach, realized by the integration of $\xi_3(t)$, turning
$\hat u$ into an estimation of the minimizer $u^*$.


\begin{figure}[htpb]
  \centering
  \includegraphics[width=\linewidth]{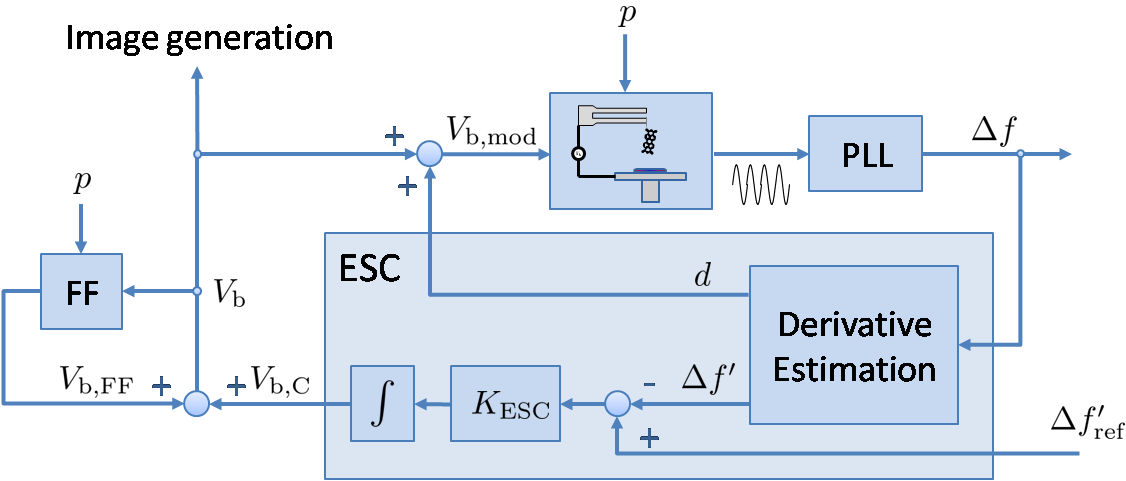}
  \caption{Block diagram of the closed loop with the ESC.}
  \label{fig:block_diagram_ESC}
\end{figure}

~\\
Applied to SQDM we have $h(u) = \Df(\Vb)$ as a local convex function and
the ESC computes the
derivative $\Df' = \frac{\diff\Df}{\diff\Vb}$ by modulating the dither
signal $d(t)$ onto the \Vb signal. 
Since $\Df' = 0$ characterizes the dips' minima, it also characterizes
exactly the value of \Vmp. 
Thus, if a potential controller regulates $\Df'$ to zero\footnote{Note
that only one dip can be tracked at a time.}, it automatically yields
$\Vb = \Vmp$.
Therefore, $\Df' = 0$ is used as the reference and the gradient descent
is achieved by using an integral controller
\begin{align*} 
  \Vbc(t) = \Kesc \int_0^t \ed(\tau) \diff\tau
\end{align*}
that minimizes the error $\ed(t) = \Dfdref - \Df'(t)$ with $\Dfdref =
0$ and $\Kesc > 0$.
The voltage applied to the AFM cantilever is
\begin{align*}
  \Vbmod(t) &= \Vbc(t) + d(t) + \Vbff(t) \ ,
\end{align*}
where \Vbff is the feedforward signal computed as detailed in
Section~\ref{sec:feedforward}.
Note that the signal $\Vb = \Vbc + \Vbff$ without the dither signal is
used for image generation.
The corresponding block diagram is depicted in
Fig.~\ref{fig:block_diagram_ESC}.

~\\
The ESC parameters that need to be chosen are the dither signal
amplitude \Ad and frequency \Wd, the low and high pass filter cut-off
frequencies \WL and \WH, and the control gain \Kesc.
In the following we present some general guidelines for choosing these
parameters based on the characteristics of the respective dip and the
\PLL.

\subsubsection{Dither Signal}
The interval in which the applied bias voltage varies locally is
$[\Vb-\Ad, \Vb+\Ad]$.
The gradient of the dip $\frac{\diff \Df}{\diff \Vb}$ is then
approximated within this interval, \ie an average gradient around the
current bias voltage \Vb is computed.
A natural upper limit is therefore the dip width $w^\mp$ itself.
On the other hand, the smaller \Ad, the more accurate the gradient at
\Vb.
However, the \Df measurements for the gradient computation are corrupted
by noise, which poses a lower limit on \Ad.
Thus, we get 
\begin{align*}
  \Ad \leq w^\mp \ ,
\end{align*}
where \Ad should be chosen as small as possible.


Regarding the choice of the dither signal frequency \Wd: the higher
the frequency, the faster the gradient estimate converges but also the higher
the variance of the estimate.
Additionally, if \Wd is much larger than the system's bandwidth, the dither
signal is damped and shifted significantly and in consequence the
gradient computation deteriorates.
Hence, an upper bound depends on the maximum bandwidth of the
system dynamics.
In SQDM, the \PLL is the limiting element with its bandwidth $\WPLL$.
We have found
\begin{align*}
  2 \WPLL \leq \Wd \leq 10 \WPLL
\end{align*}
to work well, where larger values decrease convergence time but increase
variance.

\subsubsection{Filter Cut-Off Frequencies}
The objective of the high pass filter is to remove the constant offset
of the \Df signal.
This can be sufficiently fast achieved for 
\begin{align*}
  \WH \geq 0.5\Wd \ .
\end{align*}
The objective of the low pass filter is to smoothen the signal $\xi_2$.
The smaller \WL the stronger the smoothing.
The stronger the smoothing, the less oscillatory the gradient
approximation becomes but also the slower the convergence.
Hence, choosing \WL is a trade-off between convergence speed and
variance, similar to \Wd.
We have found that 
\begin{align*}
  0.1 \Wd \leq \WL \leq 0.5 \Wd
\end{align*}
works well.
Accordingly, \WL can be chosen within this interval, depending on the
objective of fast convergence (fast scanning) or small variance.

\subsubsection{Phase Shift}
The \PLL and the high pass filter introduce an additional phase shift
\begin{align*}
  \phi = \text{arg}\big(G_\text{PLL}(\im\Wd) G_\text{HP}(\im\Wd)\big)
\end{align*}
w.r.t. the dither signal, where $G_\text{PLL}(\im\Wd) = \frac{1}{\im\Wd
+ \WPLL}$ is the PLL transfer function and $G_\text{HP}(\im\Wd) =
\frac{\im\Wd}{\im\Wd+\WH}$ the high pass filter transfer function.
This can be accounted for by adding the same phase shift $\phi$ to the
dither signal that is multiplied with the high pass outcoming signal
(see Fig.~\ref{fig:block_diagram_ESC_basic}).

\subsubsection{Control Gain}
To facilitate gain tuning, we define \Kesc via
\begin{align}
  \Kesc = \frac{k}{\abs{ G_\text{PLL}(\im\Wd) G_\text{HP}(\im\Wd) }} \ ,
  \label{eq:compGain}
\end{align}
where $k > 0$ is the new tunable ESC gain.
This redefinition automatically compensates the amplitude change of
$\xi_3(t)$ introduced by the PLL and the high pass filter.
This way the PLL and the high pass filter can be changed
without influencing the amplitude.
\green{Note that the low pass filter $G_\text{LP}(\im\Wd)$
is not included in \eqref{eq:compGain} because then one loses the
possibility of adjusting the convergence of $\xi_3(t)$ independently of
that of $\hat u(t)$.
}

The larger $k$, the faster the convergence to the minimum
but also the more oscillatory the estimated minimizer $\hat u(t) =
\Vb(t)$.
Hence, the larger $k$, the faster we can scan the sample but at the
cost of less accurate tracking.
In particular, oscillations due to high $k$ values eventually appear
in the final image as noise.
Furthermore, if the oscillations in \Vb become too large, the dip might
be lost.
For instance, when the negative dip is tracked the oscillations might cause
\Vb to leave the dip towards the left side of the dip.
There the gradient descent then leads the controller to further decrease
the \Vb value down the parabola, making it impossible to recover the
dip.

\subsection{Slope Tracking Control}
\label{sec:slope_tracking_control}

\begin{figure}[htpb]
  \centering
  \includegraphics[width=\linewidth]{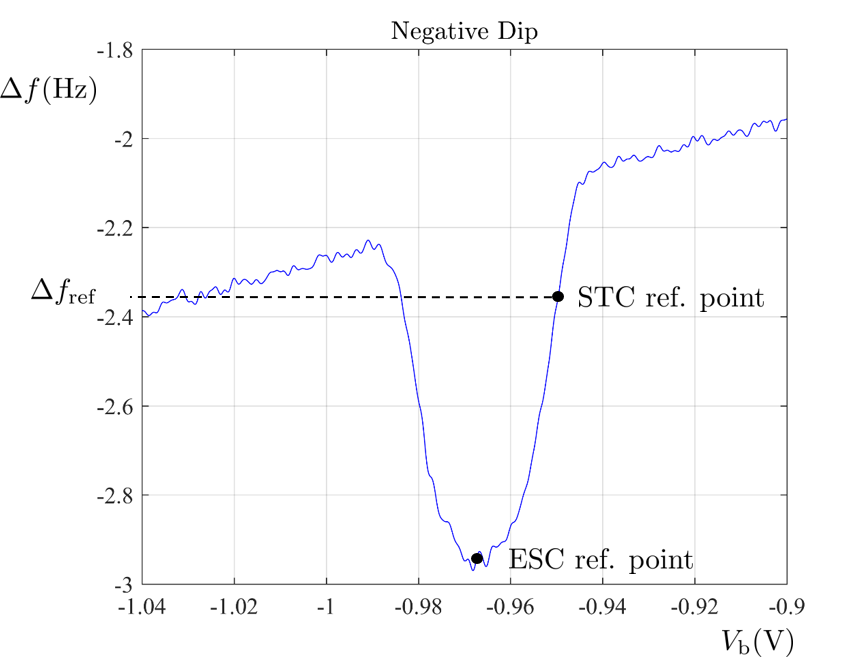}
  \caption{Measured spectrum of the negative dip with controller reference
  points for ESC and STC.
  The dashed black line indicates that the \Dfref point of the STC has
  three crossings with the blue spectrum.
  Hence, three equilibria of the closed-loop system exist for
  the STC.
  The one on the dip's inner slope (on the right hand
  side for the negative dip) is the point we want to stabilize.
  }
  \label{fig:spectrum_negDip_refPoints}
\end{figure}
\begin{figure}[htpb]
  \centering
  \includegraphics[width=0.8\linewidth]{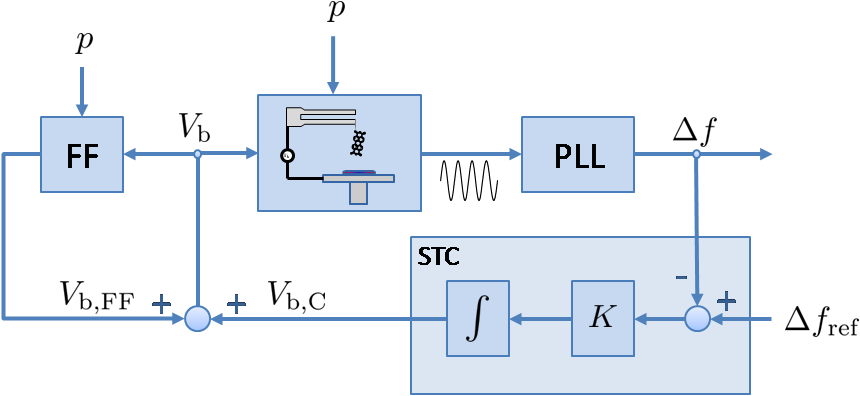}
  \caption{Block diagram of the closed loop with the STC.}
  \label{fig:block_diagram_STC}
\end{figure}

Another possibility to control SQDM is by tracking a point on the dips'
slope (Fig.~\ref{fig:spectrum_negDip_refPoints}), instead of tracking
the dips' minima.
The resulting \Df value is then set as the reference value \Dfref for
the whole sample and the deviations $\ef(t) = \Dfref - \Df(t)$
are used directly as an error to the integral controller
\begin{align} 
  \Vbc(t) = \Kstc \int_0^t \ef(\tau) \diff\tau
  \label{eq:STC}
\end{align}
that adapts \Vb accordingly with $\Kstc < 0$.
The resulting voltage applied to the AFM cantilever is
\begin{align*}
  \Vb(t) &= \Vbc(t) + \Vbff(t)
\end{align*}
with \Vbff computed as detailed in Section~\ref{sec:feedforward}.
The resulting block diagram is depicted in
Fig.~\ref{fig:block_diagram_STC}.

~\\
The STC parameters that need to be chosen are the exact position of the
reference point \Dfref and the control gain \Kstc.

\subsubsection{Reference Point}
The choice of the STC reference point \Dfref is very important.
In particular because most of the \Df values of a dip appear three times
in the local spectrum around the respective dip (see
Fig.~\ref{fig:spectrum_negDip_refPoints}).
We choose \Dfref to lie on the inner slope of the dips (\ie on the right
slope of the negative dip and on the left slope of the positive dip) and
detail the reasons in the following.

First note, that the inner slope is always larger than the outer slope
(compare Fig.~\ref{fig:spectrum} and
Fig.~\ref{fig:spectrum_negDip_refPoints}), which suggests better control
performance on the inner slope.
Second, if the reference point is located on the inner slope, according
to \eqref{eq:STC} the controller is able to drive \Vb to \Dfref even if
\Vb has left the dips towards the vertex
of the parabola (the sign of \ef doesn't change). 
%
This would not be the case if the reference point was chosen to lie on
the outer slope and \Vb had left the dips towards the part of the
parabola where values decrease indefinitely.
In that case the controller would drive the bias voltage to even larger
absolute values until the corresponding \Dfref value on the parabola is
reached (crossing of the dashed black line and the blue parabola on the
left side in Fig.~\ref{fig:spectrum_negDip_refPoints}).
In general, the STC won't be able to recover the dip in that case.

Regarding the exact position, \Dfref should lie relatively far away from the
dip minimum at \Vmp because this is another critical point for the STC.
If \Vb moves over this point (\eg $\Vb < \Vneg$ for the negative dip)
the control error \ef becomes smaller instead of larger, which
automatically has a deteriorating effect on the control performance and
increases the probability that the dip is left towards the part of the
parabola where values decrease indefinitely.

\subsubsection{Controller Gain}
\label{sec:STC_controller_gain}
Regarding the STC gain \Kstc, the same holds as for the ESC gain \Kesc.
The larger \Kstc the faster the convergence to \Dfref but also the more
oscillatory.
Thus, the larger \Kstc the faster the sample can be scanned but at the 
cost of less accurate tracking.

~\\
The STC does not require the computation of the derivative and is
therefore faster. 
However, it introduces a systematic error due to the difference between
\Vmp and the \Vb value at the STC reference \Dfref
(approximately\footnote{Since the positive dip is wider, this error is
larger for the positive dip.} $\unit[20]{mV}$ in
Fig.~\ref{fig:spectrum_negDip_refPoints}) that we denote by \eSTC.
Additionally, this error is not constant and changes while scanning
because when the dip changes its position it slides the parabola up- or
downwards.
For further clarification imagine that the negative dip moves vertically
upwards (no horizontal movement).
In that case, \Vneg does not change but $\Vb(\Dfref)$ changes to more
negative values, moving closer to \Vneg.
Hence, \eSTC decreases as the dip moves vertically upwards and decreases
as the dip moves downwards.
This effect also occurs when the dip moves along the parabola because
there is always a vertical motion component.
In particular, the effect is larger for the positive dip because it is
typically located at steeper parts of the parabola where the vertical
motion is more pronounced.

A comparison of the advantages and drawbacks of the ESC and STC is
provided in Tab.~\ref{tab:comparison_feedback}.

\begin{table}[htpb]
  \centering
  \caption{Comparison of feedback approaches 
  }
  \begin{tabular}{c|cc}
    \toprule
        & Advantages & Drawbacks \\
    \midrule
    ESC & robust     & slow, many parameters \\
    STC & fast, few parameters   & delicate, systematic error \\
    \bottomrule
  \end{tabular}
  \label{tab:comparison_feedback}
\end{table}

\subsection{Feedforward}
\label{sec:feedforward}

2DOF control approaches are well known in scanning probe techniques like
scanning tunneling or atomic force microscopy.
The control objective is usually to steer the piezo stages
that govern the movement of the microscope tip in $x$, $y$, and $z$
direction.
The $z$-piezo is controlled according to the topography feedback signal
(\eg tunneling current in scanning tunneling microscopy or force in
contact mode atomic force microscopy \cite{Schitter2003, Schitter2004}),
whereas the piezos in $x$ and $y$ direction are controlled such that the
tip follows specific reference trajectories in the $(x,y)$-plane that
implement the raster scanning pattern.
For the main scanning direction, this is usually a triangular signal
\cite{Pao2007}.
The controllers are often based on models of the respective piezo
stages.
The feedforward part of the 2DOF controllers is therefore
also usually model based and techniques like $H_\infty$, $l_1$-optimal,
model inversion, or iterative learning control are employed.
For good overviews on this topic see \cite{Pao2007, Devasia2007,
Clayton2009, Leang2009, Yong2012, Gu2016}.

As already detailed throughout this section, the objective is to track
the dips' minima (with the ESC) or a point on the dips' inner slope
(with the STC).
This objective is taken care of by the respective feedback controller.
However, as already mentioned in
Section~\ref{sec:slope_tracking_control}, it is possible (and has
occurred) that the dips change their position faster than the respective
controller can adapt the bias voltage, which eventually leads to
the controller ``losing the dip''.
A combination of a rapid change of the \ELP, a high scan speed,
and the controller dynamics is usually the cause.
In that case, the scanning process has to be aborted and restarted from
the beginning.

This risk can be substantially reduced by the generation of an
appropriate \FF signal, such that the initial value for each controller
at each tip position stays always within the dips' interval.  
In that way, the \FF has not only the potential for increased
performance, and with that eventually higher scan speeds, but is
even essential for correct operation of SQDM.

\begin{figure}[htpb]
  \centering
  \includegraphics[width=0.7\linewidth]{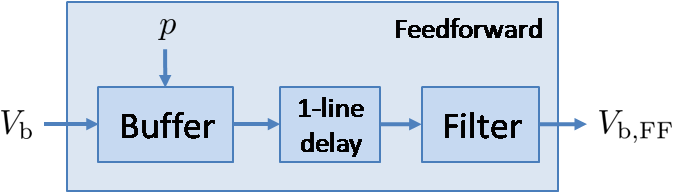}
  \caption{Feedforward block diagram. $p = (x,y)$ denotes the current tip
  position.
  Current \Vb values are stored in a buffer, delayed by one line and
  outputted through a mean filter at the next line as the \FF signal
  \Vbff.
  }
  \label{fig:block_diagram_FF}
\end{figure}

Since an a priori model of the \ELP is not available, a natural choice
is an approach that is based on the previously scanned line.
Fig.~\ref{fig:block_diagram_FF} shows a block diagram of the \FF
signal generator for SQDM, which comprises the following elements and
features.

\begin{itemize}
  \item \textbf{Buffer:}
    The $\Vb(x,y)$ values of the currently scanned line $y$ are stored in a
    buffer alongside the indexing $x$ values within the line.
    Hence, $x$ is the fast scan direction.
    At the same time, already stored $\Vb(x,y-1)$ values of the
    previously scanned line $y-1$ are used as a basis for the \FF signal
    $\Vbff(x,y)$ of the current line.
  \item \textbf{Filter:} 
    The measured and buffered \Vb values are corrupted by noise and
    small ripples caused by the ESC (if this is used for control), which
    have a deteriorating effect on the control performance,
    especially in regions where the \ELP is relatively flat.
    Therefore, while scanning the current line $y$, the previously
    measured $\Vb(x,y-1)$ values are smoothed to generate the \FF signal
    $\Vbff(x,y)$ using the mean filter
    \begin{align*}
    \Vbff(x,y) = \frac{1}{n} \sum_{i=1}^n \Vb
      \left( x-\frac{n}{2}+i, y-1 \right) 
    \end{align*}
    with filter window length $n$. 
    %
  \item \textbf{Scan speed adaptation:}
    The previous line $y-1$ is indexed using the measured $x$-position
    values.
    In the current line scan, the active $x$-position is determined and
    used for picking the right reference value. 
    This allows for varying scan speed within a line.
\end{itemize}

This approach to generate the \FF signal \Vbff is simple,
straightforward to implement, yet practically powerful.
By adding \Vbff to \Vb, the controller has only to correct the
difference between the current and the previous line.  
This results in a decreased control error and therewith a decreased 
probability that the ``dips are lost''.
Accordingly, the scan speed can be increased while maintaining the same
image quality.

\section{Simulations}
\label{sec:simulations}

\begin{figure}[htbp]
  \centering
  \includegraphics[width=1.08\linewidth]{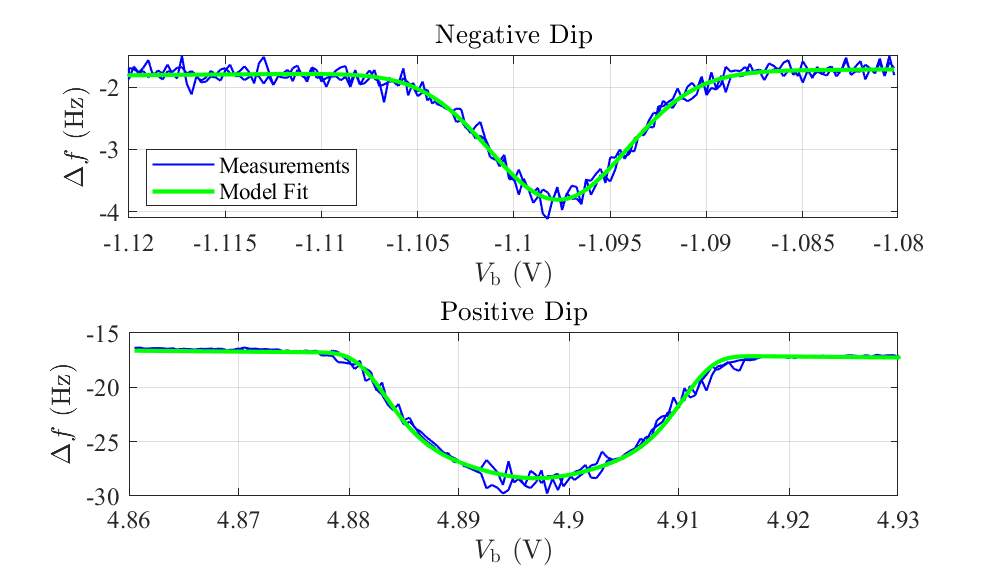}
  \caption{Experimentally acquired negative dip (top) and positive dip
  (bottom) with model fit.
  }
  \label{fig:exp_dips}
\end{figure}

In this section, we simulate the SQDM process together with the ESC and
the STC and evaluate their performances qualitatively and
quantitatively.
To this end, we use the simulation model of
Section~\ref{sec:simulation_model} implemented in MATLAB/Simulink.
We use a fixed-step \texttt{ode3} (Bogacki-Shampine) solver with a
sampling time of $T_\text{s} = \unit[5]{ms}$.
Experimentally acquired spectra (Fig.~\ref{fig:exp_dips}) and the \Vmp
maps of Fig.~\ref{fig:dipMaps} are used as reference data.
The cutoff frequency of the \PLL was determined by fitting a step
response and was computed as $\WPLL = \unit[10]{s^{-1}}$.
The normally distributed white noise has a standard deviation of
$\sigma_\text n = \unit[0.03]{Hz}$.
The parameters of the two controllers are listed in
Tab.~\ref{tab:controller_parameters}.

\begin{table}[htpb]
  \centering
  \caption{Controller Parameters (if not given otherwise) 
  }
    \begin{tabular}{c|c}
    \toprule
      Negative Dip & Positive Dip \\
    \midrule
      ESC & ESC \\
      $\Ad = \unit[1]{mV}$  & $\Ad = \unit[1]{mV}$ \\
      $\Wd = 4 \WPLL$       & $\Wd = 4 \WPLL$ \\
      $\WL = 0.2 \Wd$       & $\WL = 0.2 \Wd$ \\
      $\WH = 3 \Wd$         & $\WH = 3 \Wd$ \\
      $k = -5\cdot 10^{-5}$ & $k = -6\cdot 10^{-5}$ \\
      \\
      STC & STC \\
      $\Delta f_\text{ref} =  1 \cdot w^-$  
        & $\Delta f_\text{ref} = -0.9 \cdot w^+$ \\
      $K_\text{STC} = 0.04$  & $K_\text{STC} = -0.003$ \\
    \bottomrule
    \end{tabular}
  \label{tab:controller_parameters}
\end{table}

\subsection{Influence of the Dip Parameters}
\label{sec:influence_of_the_dip_parameters}

In Sec.~\ref{sec:control} we have established a connection, in
the form of constraints, between some of the controllers' parameters and
the parameters of the dips and the \PLL.
Here we furthermore investigate how the controllers' performance is
influenced by the dips' depth $d^\mp$ and width $w^\mp$.

\begin{figure}[htbp]
  \centering
  \includegraphics[width=1.08\linewidth]{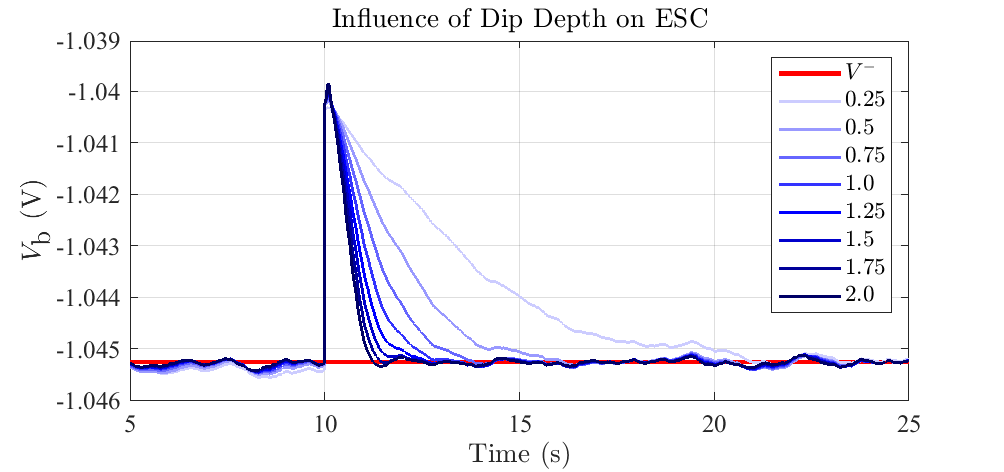}
  \caption{ESC simulations of the negative dip with gain $k = 10^{-5}$.
  The simulations were performed for different dip depths $m \cdot d^-$
  with the scaling factors $m$ as indicated in the legend.
  At $t = \unit[10]{s}$ the current bias voltage \Vb was shifted towards
  the right slope of the dip such that the ESC had to regain the minimum
  at \Vneg.
  All lines converge to the same signal because each simulation used the
  same noise realization for the sake of better comparison.
  Similar results are obtained for the positive dip.
  }
  \label{fig:influenceDipDepth}
\end{figure}

\begin{figure}[htbp]
  \centering
  \includegraphics[width=1.08\linewidth]{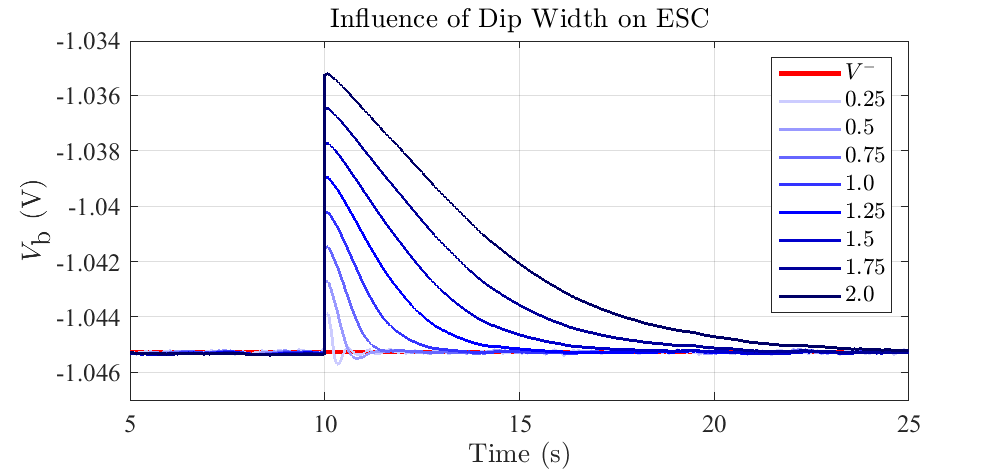}
  \caption{ESC simulations of the negative dip with gain $k = 10^{-5}$.
  The simulations were performed for different dip widths $m \cdot w^-$
  with the scaling factors $m$ as indicated in the legend.
  At $t = \unit[10]{s}$ the current bias voltage \Vb was shifted towards
  the right slope of the dip such that the ESC had to regain the minimum
  at \Vneg.
  All lines converge to the same signal because each simulation used the
  same noise realization for the sake of better comparison.
  Similar results are obtained for the positive dip.
  }
  \label{fig:influenceDipWidth}
\end{figure}

As can be seen in Fig.~\ref{fig:influenceDipDepth} and 
Fig.~\ref{fig:influenceDipWidth}, the deeper and narrower,
\ie the sharper the negative dip, the better because the faster the
convergence.
The same holds for the positive dip, as well as using the STC because
sharper dips have steeper slopes, which improve tracking.
We omit the corresponding plots for the sake of brevity.
Hence practitioners of SQDM should aim for sharp dips.
This can be achieved, for instance, with small oscillation amplitudes of
the microscope cantilever \cite{Kocic2015}.
However, at the same time such amplitudes increase the noise, which
ultimately imposes a lower bound on the oscillation amplitude.
Thus, the selection of the oscillation amplitude is a trade-off.

\subsection{Simulation Results}
\label{sec:simulation_results}

We now compare the ESC and STC with and without feedforward using 
the experimentally acquired reference data.
We start with an exemplary time evolution of the involved signals and
illustrate the influence of the feedforward.
Afterwards we turn our attention to the whole image generation and
discuss the final images qualitatively and quantitatively.
Using the same measures, we also quantify the influence of the
scan speed on the image quality.

\subsubsection{Time Evolution}

\begin{figure*}[htbp]
  \centering
  \includegraphics[width=1.08\linewidth]{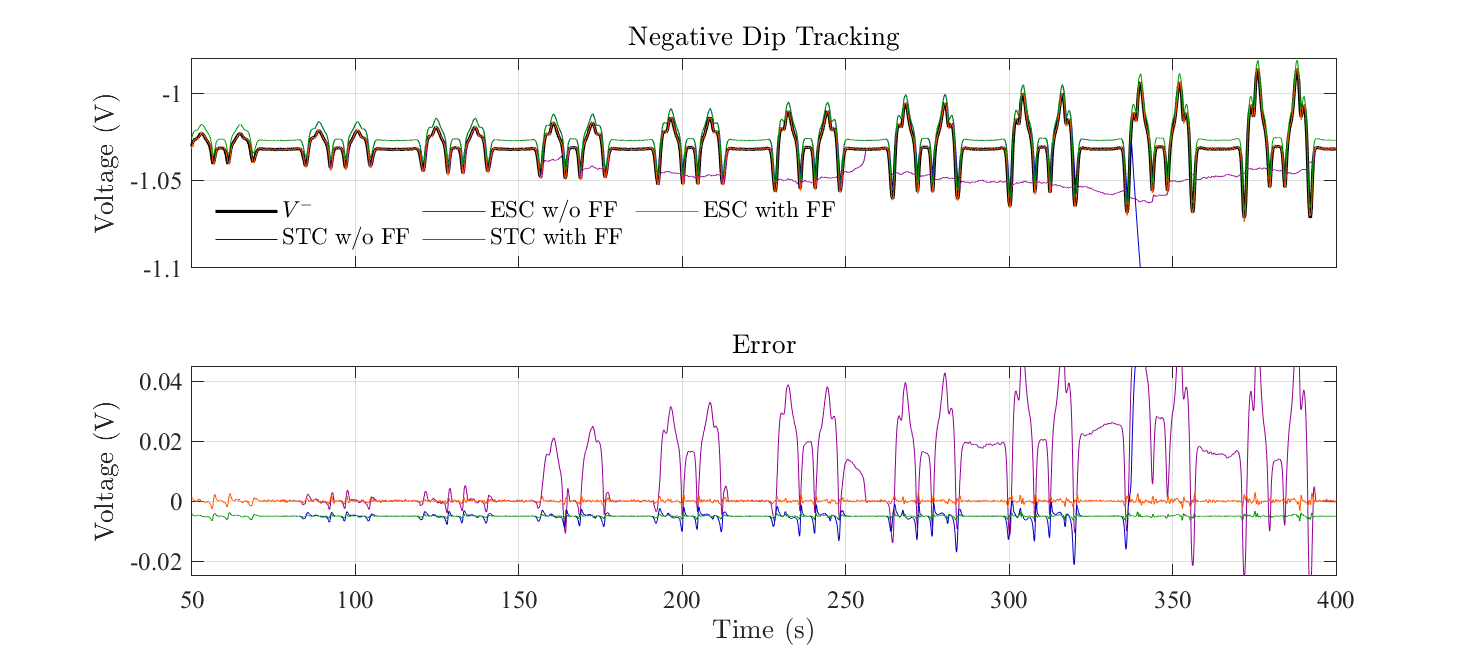}
  \caption{
    The top figure shows the reference evolution ($V^-$) together with
    the results of the different controllers. 
    The bottom figure shows the error.
  }
  \label{fig:Sim1_negDip}
\end{figure*}

\begin{figure*}[htbp]
  \centering
  \includegraphics[width=1.08\linewidth]{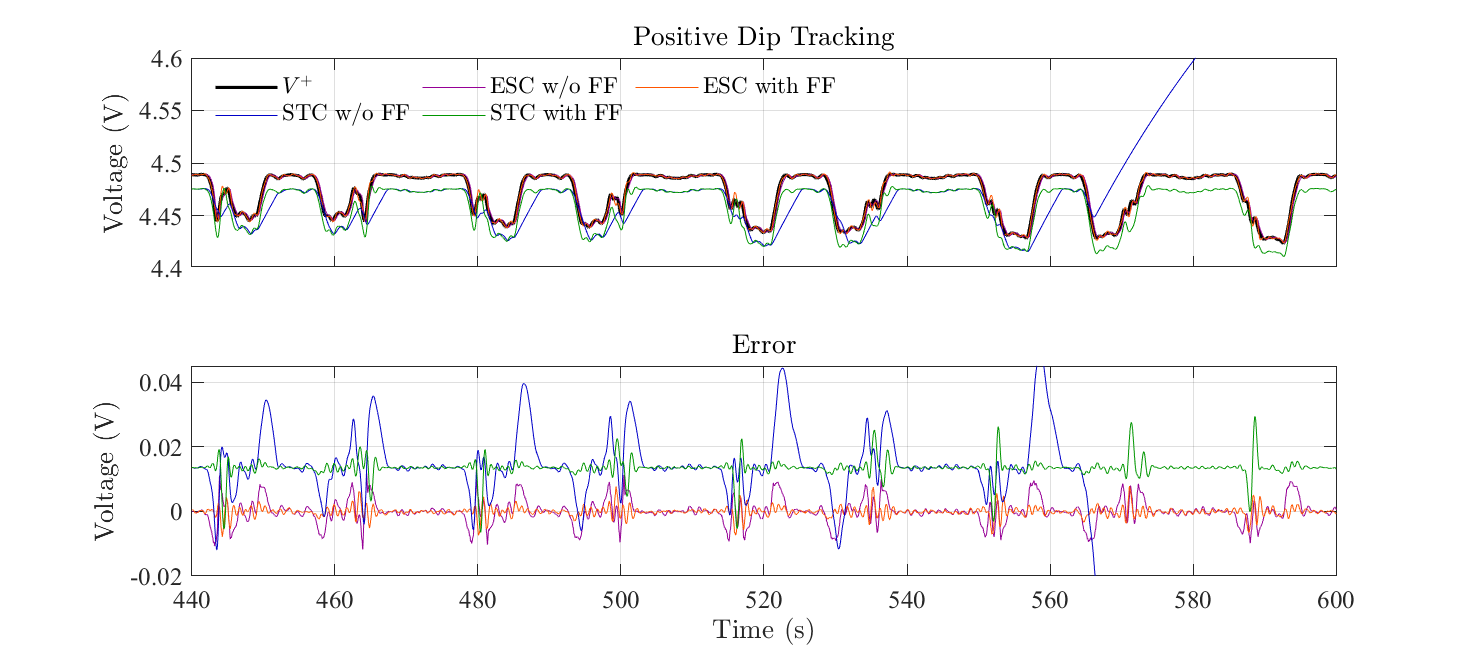}
  \caption{
    The top figure shows the reference evolution ($V^+$) together with
    the results of the different controllers. 
    The bottom figure shows the error.
  }
  \label{fig:Sim1_posDip}
\end{figure*}

In Fig.~\ref{fig:Sim1_negDip} and Fig.~\ref{fig:Sim1_posDip} exemplary
time evolutions of the \Vb signal for the ESC and STC without and with
feedforward are shown.
Both simulations were performed with a scanning time of
\red{$\Tscan = \unit[2]{h}$} for the respective dip map.
To mimic the experiment, each line is scanned back and forth.
This results in approximately \red{
$\unit[18]{s}$ for each single line
scan, which in turn equals a scanning speed of approximately
\unit[33.3]{\AA/s}
with the given area of \unit[600$\times$600]{\AA}.  
}

In case of the negative dip (Fig.~\ref{fig:Sim1_negDip}),
at the beginning of the depicted evolution all four controller instances
are able to adequately track the dip, though the STC with a constant
error due to the fact that it tracks a point on the dips' slope.
As the variations in the electric potential increase, both controllers
without the \FF are unable to fully follow the reference.
The ESC is unable to follow the $V^-$ variations
(beginning around $t = \unit[160]{s}$) and the STC loses the reference
around $t = \unit[340]{s}$.
In case of the positive dip (Fig.~\ref{fig:Sim1_posDip}),
the STC without \FF again loses the dip (around $t = \unit[565]{s}$) and
the ESC tracking results in larger errors for the shown plot as compared
to its version with \FF.
Indeed at later times, which are not depicted for reasons of clarity,
the ESC without \FF presents the same behavior as in
Fig.~\ref{fig:Sim1_negDip}, where it is unable to follow the \Vmp
variations.
Hence, we can conclude at this point that with \FF the probability of
both controllers successfully tracking the dips is significantly higher
and the resulting tracking error is reduced.

\subsubsection{Image Plots}

Now we turn to the simulation of the complete scans of the negative
and positive dip maps $\Vmp(x,y)$ and the resulting \ELP $\Pots(x,y)$
computed by \eqref{eq:SQDM}.
For the sake of brevity we provide only the final $\Pots(x,y)$ images.

\begin{figure}[htpb]
  \centering
  \includegraphics[width=\linewidth]{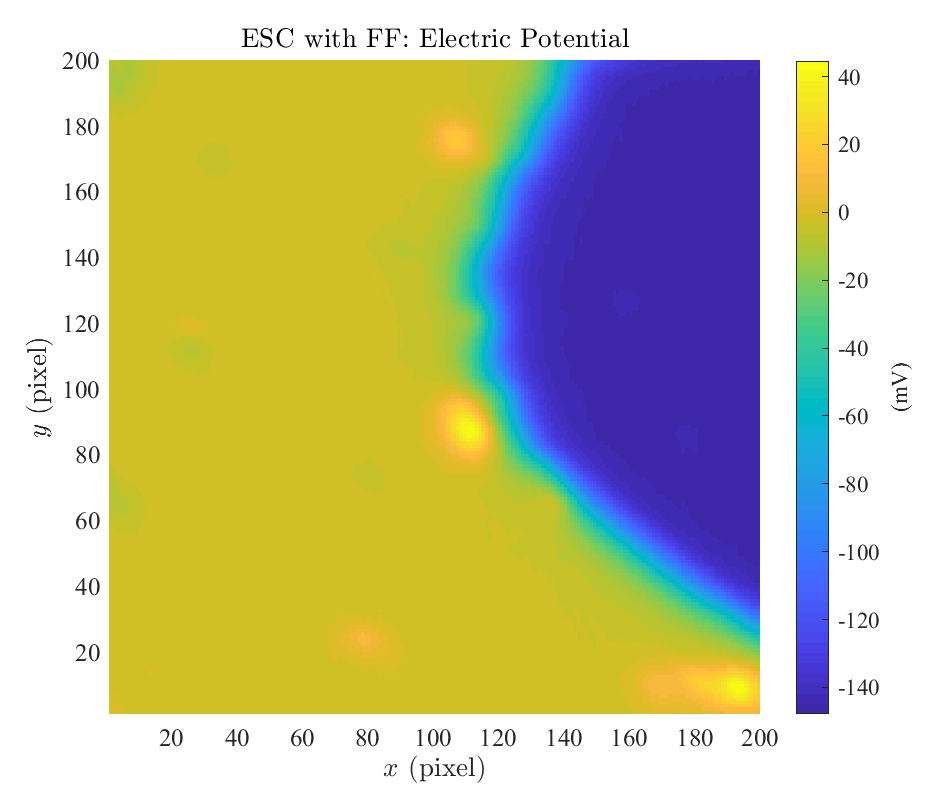}
  \caption{Resulting image with the 2DOF controller employing the ESC.}
  \label{fig:image_esc_FF_elPot}
\end{figure}

\begin{figure}[htpb]
  \centering
  \includegraphics[width=\linewidth]{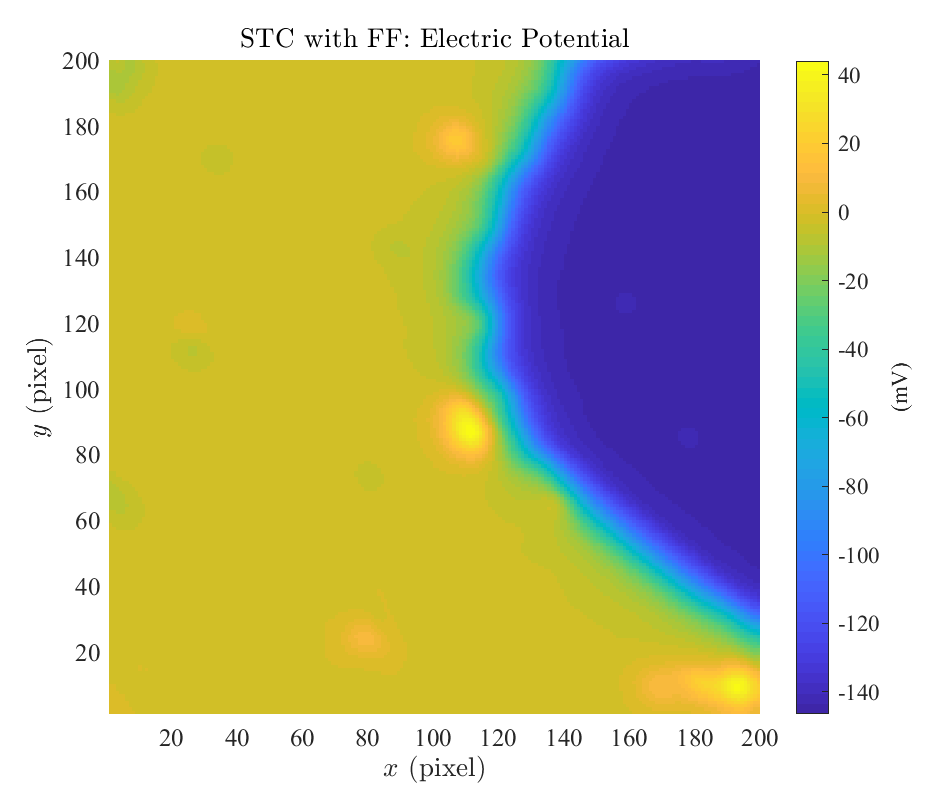}
  \caption{Resulting image with the 2DOF controller employing the STC.}
  \label{fig:image_stc_FF_elPot}
\end{figure}

\begin{figure}[htpb]
  \centering
  \includegraphics[width=\linewidth]{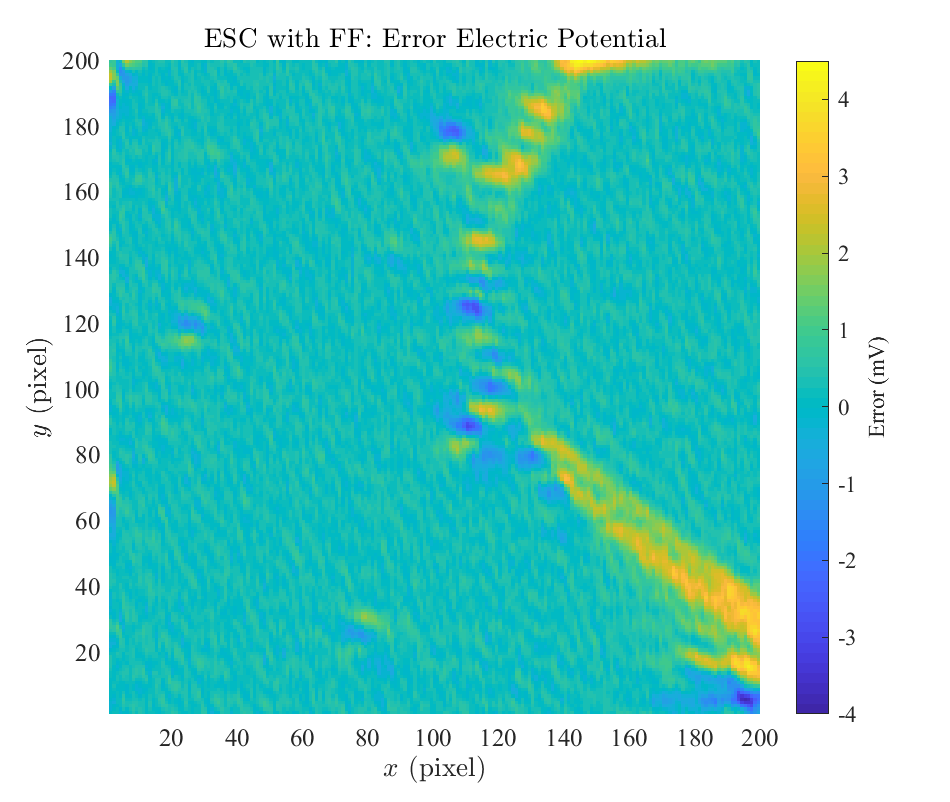}
  \caption{Resulting image error with the 2DOF controller employing the ESC.}
  \label{fig:image_esc_FF_elPot_error}
\end{figure}

\begin{figure}[htpb]
  \centering
  \includegraphics[width=\linewidth]{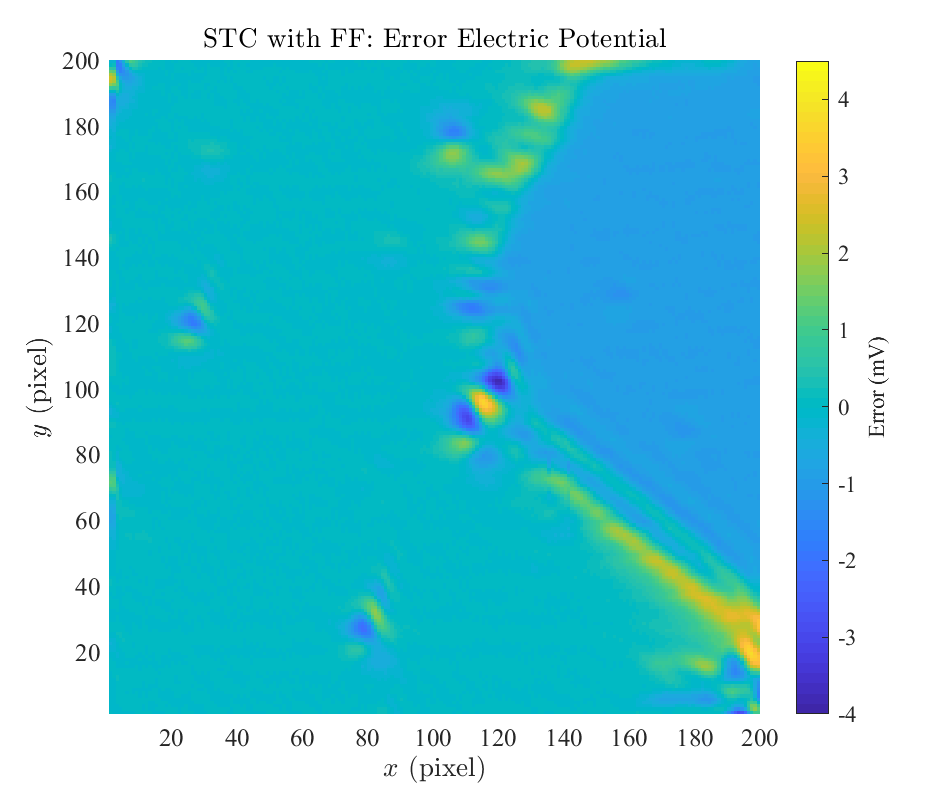}
  \caption{Resulting image error with the 2DOF controller employing the STC.}
  \label{fig:image_stc_FF_elPot_error}
\end{figure}

The two resulting potential images of the ESC+FF
(Fig.~\ref{fig:image_esc_FF_elPot}) and STC+FF
(Fig.~\ref{fig:image_stc_FF_elPot}) look almost identical to the
reference in Fig.~\ref{fig:STM_SQDM} and \ref{fig:dipMaps}.
Hence, the dips are successfully tracked over the whole sample.
The respective error images (Fig.~\ref{fig:image_esc_FF_elPot_error}
and Fig.~\ref{fig:image_stc_FF_elPot_error}) reveal that the errors are 
in the range of \red{$\unit[-4]{mV}$ to $\unit[4.5]{mV}$}.
As the total \Pots variation is $\unit[190.5]{mV}$, the relative errors
w.r.t. this variation are \red{$\unit[-2.1]{\%}$ to $\unit[2.4]{\%}$}.
Furthermore, it can be seen that the STC version of the 2DOF controller
performs slightly better with smaller errors on average and with less
noise.
On the other hand, above the molecular island (right part in
Fig.~\ref{fig:image_stc_FF_elPot_error}) it can be observed that the
average STC error is shifted roughly about \unit[1.5]{mV} towards more
negative values.
This is due to the change of the systematic STC error \eSTC as discussed
in Section~\ref{sec:STC_controller_gain}.
Nevertheless, the STC performs slightly better than the ESC, as is also
confirmed in Tab.~\ref{tab:MSE_PSNR}, where we quantify the image
quality using the image mean-square error (MSE) and the image peak
signal-to-noise ratio (PSNR) per pixel respectively\footnote{These
measures are often used in imaging science as objective quality
measures. See, for instance, \cite{Avcibas2002, Sheikh2006}.}.

The total scan time for each of the results in
Fig.~\ref{fig:image_esc_FF_elPot} and Fig.~\ref{fig:image_stc_FF_elPot}
was \unit[4]{h}.
This is approximately 17 times faster than the original image generation
process based on spectroscopy grids that would have taken
\unit[66.7]{h} (see Sec.~\ref{sec:original_image_generation_process}).




\begin{table}[htpb]
  \centering
  \caption{MSE and PSNR per pixel for Fig.~\ref{fig:image_esc_FF_elPot} and
    Fig.~\ref{fig:image_stc_FF_elPot}. 
    }
    \begin{tabular}{c|c|c}
    \toprule
        & MSE       & PSNR  \\
        & $[$mV$]$  & $[$dB$]$ \\
    \midrule
      ESC & 0.427 & 63.7 \\
      STC & 0.346 & 64.6 \\
      \bottomrule
    \end{tabular}
  \label{tab:MSE_PSNR}
\end{table}

\subsubsection{Influence of scan time}

\begin{figure}[htpb]
  \centering
  \includegraphics[width=\linewidth]{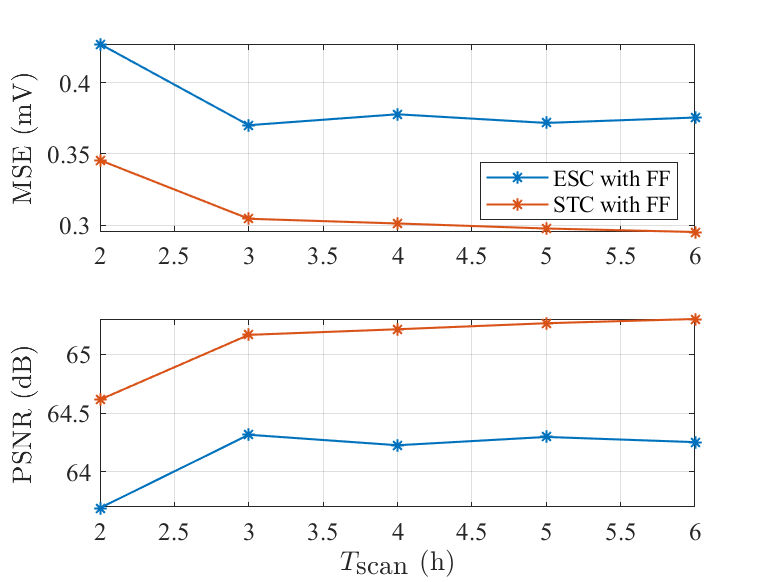}
  \caption{Influence of increasing scan times on the mean square error
  (MSE) and the peak signal-to-noise ratio (PSNR) of the resulting \ELP
  image.}
  \label{fig:image_MSE_PSNR}
\end{figure}

Finally we investigate how the resulting image quality is influenced by
the scan time \Tscan.
To this end we repeat the simulation with scan times from
\red{\unit[2]{h} to \unit[6]{h}.
This equals scan speeds of \unit[33.3]{\AA/s} to \unit[11.1]{\AA/s}.}
After every simulation we compute the MSE and PSNR of the resulting
\ELP image and plot it over \Tscan.
The results are depicted Fig.~\ref{fig:image_MSE_PSNR}.
The slower we scan, the slower the \Vmp variations
and thus, \red{the feedback controllers are better able to follow the
references, which is reflected in lower MSE and larger PSNR values}.

~\\
\emph{Remark:}
Of course, if the variability in the \ELP is large enough, there exists
always a scan time that is too small such that good tracking is not
possible, even with the \FF.
Nevertheless, we could show in this section that the probability to
successfully track the reference is significantly higher if the proposed
\FF signal is employed and that reasonable scan times can be reached for
large samples.

\section{Experimental Results}
\label{sec:experimental_results}

In this section, we present information on the implementation of the
2DOF controller in the experimental setup, discuss the SQDM operation
procedure, and present further experimentally generated images of
nanoscale samples with the presented control approach of this work.

\subsection{Implementation}
\label{sec:implementation}

The 2DOF controller is built in MATLAB/Simulink and automatically
converted into C-code and then loaded to and executed on a
ds1104 controller board by dSPACE.
The controller board is mounted into a standard desktop PC and connected
via analog-digital and digital-analog converters to a
Createc non-contact atomic force/scanning tunneling microscope
(STM/NC-AFM) that operates at $\unit[5]{K}$ and under ultra high
vacuum.
The microscope is equipped with a qPlus sensor (\cite{Giessibl2004})
tuning fork with resonance frequency $f_0 = \unit[31.2]{kHz}$, stiffness
$\kappa_0 = \unit[1800]{Nm^{-1}}$.
An amplitude of $A = \unit[0.2]{\AA}$ was used in the measurements.

\subsection{Image Generation Procedure}
\label{sec:procedure}

The procedure used to acquire a new image is as follows:
\begin{enumerate}
  \item Move to first pixel and measure the local spectrum of the
    respective dip $\Delta f^\mp(\Vb)$.
  \item Adjust \Vb manually to the reference point (the selection of the
    reference point depends on whether the STC or the ESC is used, see
    Fig.~\ref{fig:spectrum_negDip_refPoints}).
  \item Start the 2DOF controller.
  \item Start the raster scanning protocol.
  \item Enable the \FF after one or more lines have been scanned.
  \item Increase scanning speed desired.
  \item Finish the scan for the current dip $\Delta f^\mp(\Vb)$.
  \item Goto 1) and repeat for the other dip $\Delta f^\pm(\Vb)$.
\end{enumerate}

\subsection{Images}
\label{sec:images}

\begin{figure}[tb]
  \centering
  \includegraphics[width=\linewidth]{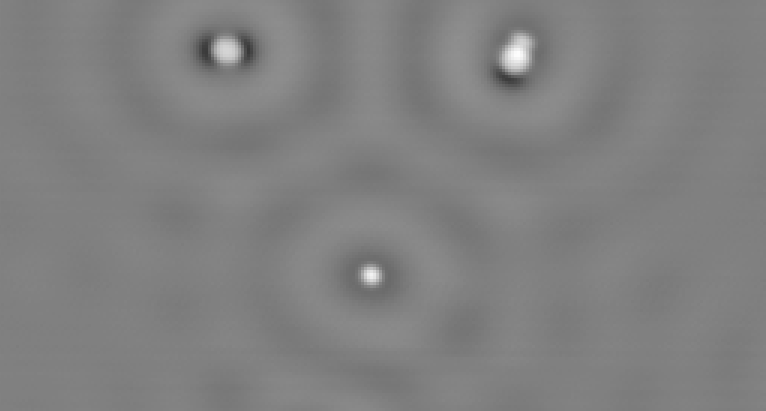}
  \caption{Scanning tunneling microscope image of a
  sample with three distinct features.
  Measured with a voltage of \unit[20]{mv} and a tunneling current of
  \unit[50]{pA}.
  }
  \label{fig:experiment_STM_image}
\end{figure}

\begin{figure}[tb]
  \centering
  \includegraphics[width=\linewidth]{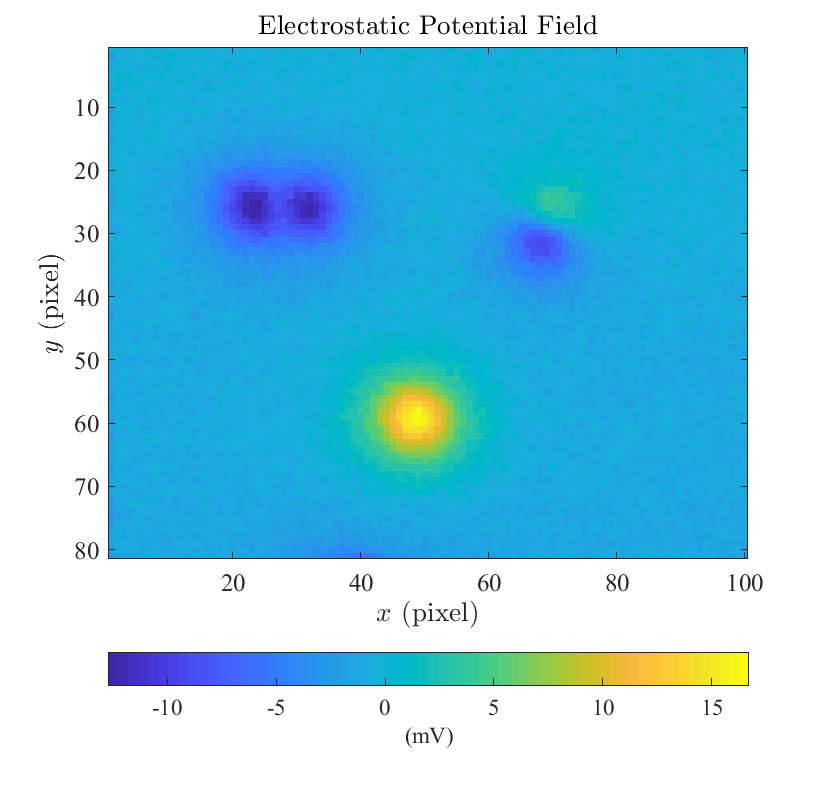}
  \caption{SQDM image of the \ELP \Pots of the three features in
  Fig.~\ref{fig:experiment_STM_image}.
  The image \unit[260$\times$210]{\AA} was divided into
  \unit[100$\times$81]{pixels} pixels and measured at a height of $z =
  \unit[22]{\AA}$ above the surface.
  The scan speed was \unitfrac[14]{\AA}{s}, resulting in a scan time of
  \unit[50]{min} per dip.
  }
  \label{fig:experiment_SDQM_image}
\end{figure}

Here we show a so far unpublished result obtained with the described
control approach.
The 2DOF controller with the STC has been employed together with the
above described procedure to investigate the \ELP of the sample
presented in Fig.~\ref{fig:experiment_STM_image}.
The resulting SQDM image is shown in
Fig.~\ref{fig:experiment_SDQM_image}.
The sample shows three features on a Ag(111) surface, namely a
PTCDA molecule (top left), a PTCDA-Ag$_2$ complex (top right),
and a single Ag adatom (bottom).

\section{Conclusion}
\label{sec:conclusion}

In this paper we have presented a two-degree-of-freedom control approach
for \SQDM. 
The approach consists of a feedback part and a feedforward signal
generator, where the latter is based on the previous line scan.
For the feedback part we have presented two different controllers,
namely an extremum seeking control approach that directly tracks the dip
minimum and a controller that tracks a reference point on the dip slope.
We have discussed the individual working principles, respective
advantages and drawbacks, and provided guidelines for controller
parameterization tailored to \SQDM.

In simulations we could show how the utilization of the \FF decreases
the probability of losing the dips and decreases the resulting
image error at the same time.
This leads to a several times faster image generation process and
enables to scan larger images in reasonable time than before, which was
also verified in experiments.
In addition, the presented control approach now allows to continuously
scan a sample, which puts \SQDM in line with other scanning
probe microscopy techniques like scanning tunneling or atomic force
microscopy.

Furthermore, we could also show that the sharpness of
the dips plays a central role in the control performance, which led to
the recommendation that one should aim for sharper dips if possible.


The next steps are going to include an automatic gain adaptation for
varying scan speeds and the improvement of the feedforward by using
previous lines to generate an accurate prediction of the current line
using a Gaussian process and implement it in the experimental setup
\cite{Maiworm2018}.

\bibliographystyle{ieeetr} 
\bibliography{bibliography}

\end{document}